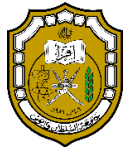

Sultan Qaboos University
College of Economics and Political Science
Information Systems Department
Data Analytics for Business

Case study

# Vanderbilt University Medical Center Data Chaos

Authors:
Azza Al Aghbari 93329
Zinah Al Maskari 135973

## Table of Contents

**Introduction**

The healthcare industry is growing rapidly in the United States because of the increased number of the aging population, shared consciousness of personal health problems, and medical technology improvements. As a result of the growing industry of healthcare, new emerging issues occur in the collection and storage of patient data, and new ways to process, analyze and distribute these data. This has exposed various security threats to personal health data (Lee, 2022). As (Al-Harrasi et al., 2021) remarked "Data are considered to be intellectual assets in organisations worldwide. Data theft is a major part of the insider threat and, for this reason, the prevention of data theft and data leakage from insider threats is becoming a major area of interest within the field".

In this case study, we will discuss the issues Vanderbilt University Medical Center (VUMC) challenges while implementing EHR systems which are used to analyze and monitor health records by the users such as doctors, organizations staff, and pharmaceutical agencies(Kaul et al., 2020), and we will analyze these issues and provide solutions and recommendations to solve them.

In 1997, VUMC started gathering data as part of its EHR initiatives. The center required greater, more structured data handling by 2009. Hospital executives started an initiative to create a data governance infrastructure at that time. Putting Data Governance into Practice. The executive staff at VUMC had various challenges:

- Although IT investments and technologies were constantly growing, HIM (Healthcare Information and Management) policies did not apply to them. All data are not created equal, and using technology by itself won't help patients receive better care. Providers and organizations must be able to tell the difference between a surplus of data, useful data, and data integration. Whereas new technology and treatment modalities are changing and expanding at a rapid rate, healthcare companies are challenged to face these data issues in their daily operations and workflow.
- Medical records became increasingly susceptible to hacking, as they were made electronically so they could be shared and transmitted easily. Cybercriminals are drawn to healthcare data because it contains financial and personal information, can be used as leverage in extortion, and—most lucrative of all—is perfect for fraudulent invoicing. Due to the fluid and constantly changing nature of a patient's medical care as well as the sheer volume of physicians, facilities, and transactions necessary to connect patient care across various settings, they are also incredibly susceptible to penetration.
- Keeping up with the emergence of new electronic information applications was challenging for the medical facility. Were, (Curioso et al., 2022) "The health industry faces significant obstacles in resource-constrained environments, including effective health innovation projects and health management programs. Additionally, one pertinent difficulty is enhancing health system operations to accomplish health sector goals. We risk digitizing chaos if procedures are not mapped, updated, or reviewed on a regular basis."

**Identifying issues**

In the healthcare industry, many hospitals are facing similar issues in implementing separate applications for each department without any integration between them, this creates a big gap between the departments patient data and it needs more effort to solve such problems. When healthcare organizations want to develop one integrated system that can hold all the health records electronically by implementing EHR, there will be three main issues derived from this problem: faulty data, data breaches, and the cost of failure.

### 1. Faulty Data

Data will not be so accurate or complete in EHR systems because of improper lab and imaging results or medical errors which are mistaken by the physician documentation, these will damage patients data and the accreditation of organizations. As (Li et al., 2016) noted, Patients data are considered as big data and it is a challenge to merge all data from each department for each patient, and this can cause fault data because medical records can be different from one database to another. Because of transmitting and recording some data with errors, crashes on devices or malicious purposes can result in exploited, outdated or missing data. Wrong diagnoses can result from the invalid measurements of patients information which will cause some significant consequences, also financial losses can occur because of faulty, untrusted data and payment errors, which will affect the organizations revenue.

### 2. Data Breaches

A huge number of people are affected (which exceeds 25 million persons) due to the data breaches that occurred from digital transformation of HIS (healthcare information systems). These data breaches come up from the data stolen from laptops, or any portable devices, these breaches will destroy the trust of patients and are very expensive to retrieve. As (Lee, 2022) commented, new advanced techniques are rapidly increasing because of cybercriminals attacks which makes it more difficult to protect patients' data, such attacks occur because of outdated systems, cyber staff insufficient experience, and extremely valuable data which has incentives to payoff these data. Spread devices of medical IoT (Internet of Things) are rising new threat to secure healthcare organizations data (Lee, 2022). Moreover, money laundering activities may occur in healthcare systems and this may affect the patients trust, as (Shaikh & Nazir, 2020) noted "Money laundering is one of the complex financial crimes that abuse an effective operating international financial system. It is structured crime which is growing day-by-day with new tricks and techniques. It has increasingly became a serious concern as it promotes terrorism, and therefore serious preventions must be enforced with the use of advanced technologies".

### 3. Cost Failure

Cost is one of the biggest barriers to EHR integration that healthcare organizations face. Nevertheless, it is shown to be a wise investment; a well-designed system increases profits, lowers expenses, and boosts productivity. EHR failures can take many different shapes, affect a wide range of areas, and frequently involve failures in several of these areas.

- **Risk to patient safety**: Doctors, nurses, and other clinicians may have workflow issues or inefficiencies as a result of the design, customization, and usage of electronic health records (EHRs), which may fail to stop patient harm or maybe contribute to it. For instance, a doctor can order the incorrect medication for a patient as a result of a medication list that is unclear. Clinical decisions based on laboratory test findings that aren't time and date-stamped risk using out-of-current information. Medical errors could also result from systems' inability to send alarms regarding dangerous prescription combinations, which can happen as a result of facility modifications, changes in how physicians enter data, or problems with EHR design. (The Pew Charitable Trusts et al., 2018)
- **Ineffective system use and poor system usability**: Errors can arise from incorrect system use in addition to EHR design elements and functions that may potentially contribute to poor healthcare quality. Usability issues are brought on by complex systems, a lack of user-friendly features (such as confusing user interfaces), incompatible workflows, or user limits. When there is a confusing screen display or when wrong data come from code that incorrectly transforms from one measuring system to another, faulty functionality could lead clinicians astray. (Bowman, 2013)

Both factors are affecting the organizations budget and decreasing ROI, increasing the users efforts and consuming their time to surpass these challenges.

**Analysis of issues**

1. **Faulty Data**

In 2019, Kaiser Family Foundation Health Tracking Poll conducted a random survey among 1190 adults (Kaiser Family Foundation, n.d.) from the national population who are living in the United States on how they experienced the EHR system and what are their attitudes about it. (Munana et al., 2019)

They found that 67% of patients noted that there is no error when doctors use EHR, 6% noted that doctors do not use EHR, and 21% of patients noticed an error when

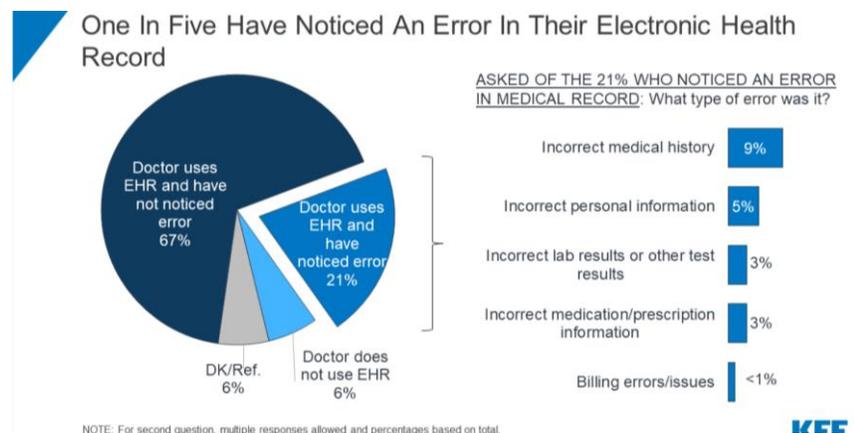

*Figure 1: One in five have noticed an error in their EHR*

doctors used EHR, the majority noted that their medical history is incorrect which was 9%. So if we want to find how many patients have recognized errors in their EHR:

$1190 \times \frac{21}{100} = 249.9$, approximately 250 patients.

If we assume that each one from the U.S. population is registered in EHR system and we want to calculate the approximate number of the population who might face these errors in their EHR in the year 2019:

$328{,}239{,}523 \times \frac{21}{100} = 68{,}930{,}300$

Up to 30 million patients might have an incorrect medical history, and about 3 million will have billing errors. This is a huge number to detect and it is affecting the healthcare industry in a negative way, patients may lose the trust on hospitals system because of the wrong patient history and personal data, and it can also cause some medical diagnosis faults as a result of the incorrect lab results and medication, and it will affect the hospitals financials because of the billing issues.

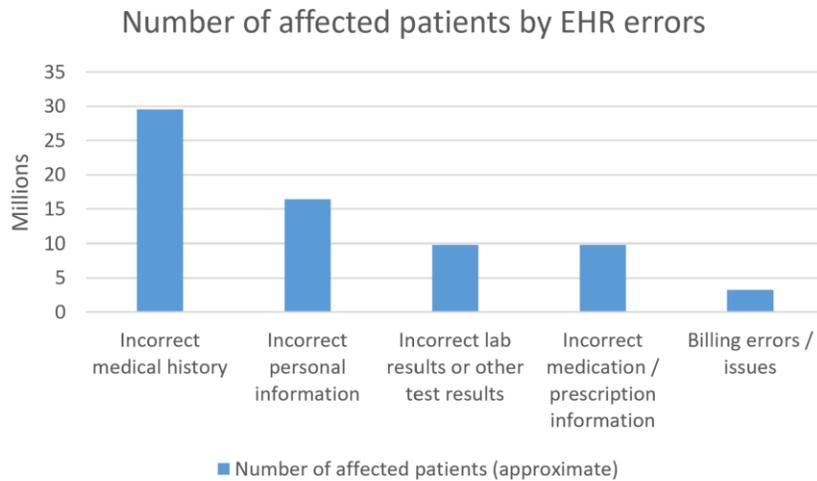

Figure 2: Number of affected patients by EHR errors

## 2. Data Breaches

According to the Healthcare Breach Report released in 2019 by Bitglass, analyzed collected data from the Department of HHS (Health and Human Services) in the U.S. about PHI (protected health information) breaches. These breaches are categorized into four types:

- Hacking or IT Incidents
- Unauthorized Access or Disclosure
- Loss or Theft
- Other (e.g. inappropriate disposal of data)

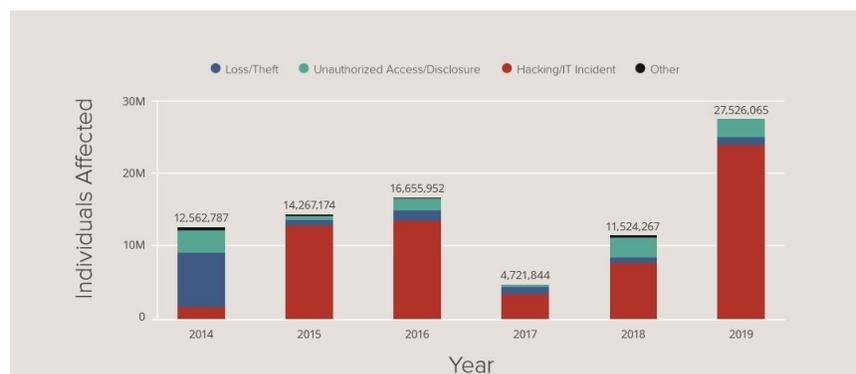

Figure 3: Individuals affected year over year

The number of individuals who are affected by data breaches in 2018 doubled in 2019 and reached approximately 27.5 million. Hacking and IT incidents were the major issue in healthcare data breaches in most of the years, although, in 2019 approximately 24 million individuals were affected by IT incidents and hacking which we can measure as 87.3% of total breaches. This takes the top because healthcare data is sensitive and it is targeted more by cybercriminals. On the other hand, the number of stolen and lost devices dropped over the years to reach approximately 1 million in 2019.

The statistics recorded that each breach cost $429, so we can calculate the total data breaches cost in healthcare organizations: 27.5 million (records) × $429 = $11.8 billion. (Bitglass, n.d.) Organizations are losing a massive amount of money from these breaches.

### 3. Risk to patient safety

Comparative Benchmarking System (CBS). The CBS database, which represents more than 400 hospitals and healthcare institutions, contains approximately 400,000 malpractice lawsuits as well as open and closed claims from captive and commercial insurers. cases chosen between January 1st, 2011, and December 31st, 2015. Open and closed malpractice lawsuits and claims with one or more EHR identifiers as relevant causes were used in this analysis. (Greenberg & Ruoff, n.d.)

According to the study's results, there are adverse events connected with using electronic medical record systems, they are common across the spectrum of healthcare settings, They are associated with a significant risk of serious damage and death, and they involve both technological and human errors .The top five causes of EHR-related instances in this survey, which collectively account for 50% of the cases, are user error, incorrect information in the record, pre-populating (copy/paste errors), conversion issues, and system issues.

**EHR-related adverse events involve both user- and system-related issues.**

EHR-related Factors Contributing to Patient Harm

| TOP FACTORS | % CASES* |
| --- | --- |
| user error | 17% |
| incorrect information in record | 16% |
| pre-populating or copy/paste errors | 14% |
| conversion issues (hybrid paper & electronic records) | 13% |
| system/software design issues | 12% |

*a case may have more than one error identified

N=420 MPL cases asserted 1/1/11–12/31/15 with an EHR-related factor identified

*Figure 4: EHR- related adverse events involve both user and system related issues*

### 4. Ineffective system use and poor system usability

Patient safety reports from 2013 to 2016 were analyzed. The Pennsylvania Patient Safety Authority database, which collects reports from 571 healthcare facilities in Pennsylvania, had the reports that were obtained. As noted in figure 5 (truenorth, n.d.), there were six categories of usability, the first challenge was data entry (27%, n = 152), given that the clinicians' workflow makes it difficult or impossible for them to enter needed information into the EHR in the proper manner. The second challenge was alerting (22%, n = 122) EHR notifications or other feedback that are insufficient

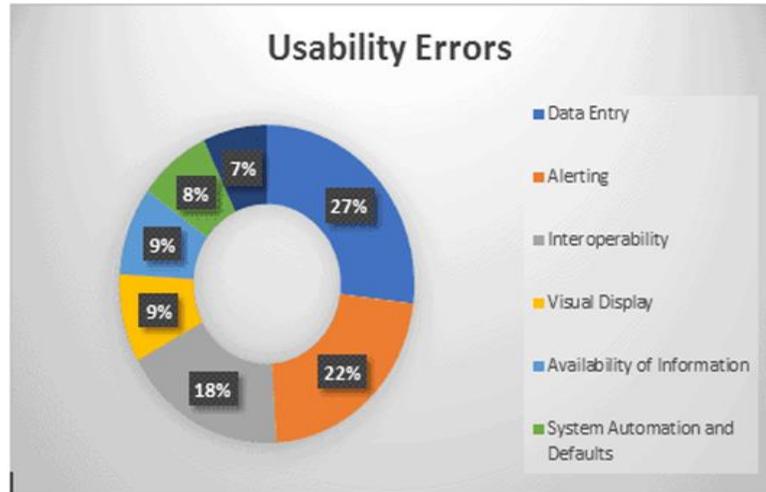

Figure 5: Usability errors

because they are missing, inaccurate, or unclear. The third challenge was interoperability (18%, n = 102) the interoperability of EHRs is insufficient within elements from the same EHR or transferred from the EHR to other systems, limiting communication between information. The fourth challenge was visual display (9%, n = 52) Clinicians have trouble evaluating information because of the unclear, crowded, or inaccurate information displayed in the EHR. The fifth challenge was the availability of information (9%, n = 50) clinically important information is not easily accessible in EHRs because it is entered incorrectly, kept incorrectly, or is otherwise unavailable. The sixth challenge was system automation, and defaults (8%, n = 43) the EHR automates or uses information that is unexpected, unexpected, or unclear to the physician. EHR usability might have had a role in a few potential patient harm incidents. Because safety reports only include a tiny portion of actual safety occurrences, the study was conservative and only a small percentage of possible injury events were linked to EHR usability. (Keene et al., 2018)

**Solutions and Recommendations**

As a result of these significant challenges, there are some approaches used to minimize categorical data conflicts:

### 1. Ensuring the source reliability to resolve conflicts by voting

One of the major approaches used and studied for years is conducting a vote for specific information and the one which got the highest occurrence number is considered as the right response, and we can calculate the mean or median for the continuous values as the right answer.

Even if this approach is valid but it has some limitations, for example, if we take the average of values it means that all values are weighted equally but in real life they are not equal. Also, if there are some existing low-quality information source providers, it will be hard to estimate the correct source reliability, such as improper sensors that release incorrect data and the false data spread by spam users. (Li et al., 2016)

## 2. Truth discovery approaches to resolve the heterogeneous data which are influenced by insufficient knowledge of source reliability

Many approaches are proposed to discover true facts and they estimate the information well. Although, these approaches may not work appropriately for combined inference heterogeneous data and these data are present everywhere. So, during the estimation of source reliability, we should measure how the input is close to the right answer and category. (Li et al., 2016)

## 3. Using HIMSS stage 7 to increase the efficiency of EHR systems, fostering internal change in the organization, boosting outcomes improvements, and realizing IT Return on Investment (ROI).

HIMSS stage 7 is providing significant solutions to overcome healthcare challenges, increasing the quality of patient care and data safety. Using HIMSS stage 7 also has benefits to support the healthcare strategy (Sulkers et al., 2018),  so healthcare facilities and health IT developers can identify safety issues at all stages of a product's development, including product design, submission to the government for clarification, deployment in hospitals and other institutions, staff training, and use. It also minimizes EHR safety issues because of the fundamental design of a product, during installation, deployment, or customizations, and of providing special workflows inside a facility. For example, the primary consumers of EHRs are physicians, nurses, and other healthcare professionals. They utilize these technologies on a daily basis, so they can spot flaws or possible issues and will help an organization develop a culture by providing feedback and reporting to internal quality and safety executives.

HIMSS are experts in data management and lead the performance capability system across the healthcare industry(Low et al., 2019). It is supporting healthcare organizations by reducing the restriction of use, increasing the compatibility of health clinical recommendations, minimizing the infection outbreak number, and saving costs resulting from substantial decision-making. (Sulkers et al., 2018)

## 4. Artificial intelligence

(AI)-based technologies that enhance workflows and increase accessibility to patient data and problem-specific information at the point of care which can be used to increase EHR usability. In order to improve patient care, reduce annoyance, and prevent workflow breakdown, businesses can deploy systems that rapidly recognize and evaluate clinical data from various reference points using AI technologies. (Chetu, n.d.)

**Conclusion**

As a result of adopting different programs for each department without any connection, many hospitals in the healthcare sector are struggling with the same challenges. This causes a significant gap in the departments' patient data, and resolving these issues requires additional work, such problems are faulty data, data breaches and risk to patient safety. So Electronic health records (EHR) are essential to providing safer treatment since they give quick access to coordinated medical records, decision assistance, clinical warnings, and follow-up/screening reminders. However, they are not without limitations. Identifying the causes of these issues, analyzing the problems, and suggesting solutions, will develop the EHR by generating a new versions and following the recommendation strategies which can create significant secure systems. Significant solutions are being offered by HIMSS stage 7 to address healthcare issues, hence improving patient care and data security.

# References

Al-Harrasi, A., Shaikh, A. K., & Al-Badi, A. (2021). Towards protecting organisations' data by preventing data theft by malicious insiders. *International Journal of Organizational Analysis*. https://doi.org/10.1108/IJOA-01-2021-2598

Bitglass. (n.d.). *Bitglass 2020 Healthcare Breach Report: Over 27 Million People Affected in Healthcare Data Breaches Last Year | Business Wire*. Retrieved November 14, 2022, from https://www.businesswire.com/news/home/20200219005071/en/Bitglass-2020-Healthcare-Breach-Report-Over-27-Million-People-Affected-in-Healthcare-Data-Breaches-Last-Year

Bowman, S. (2013). Impact of Electronic Health Record Systems on Information Integrity: Quality and Safety Implications. *Perspectives in Health Information Management*, *10*(Fall). /pmc/articles/PMC3797550/

Chetu. (n.d.). *5 Common EHR Implementation Challenges | Chetu*. Retrieved November 14, 2022, from https://www.chetu.com/blogs/healthcare/ehr-implementation-challenges.php

Curioso, W. H., Ting, D. S. W., Ginneken, B. van, & Were, M. C. (2022). Challenges in digital medicine applications in under-resourced settings. *Nature Communications 2022 13:1*, *13*(1), 1–5. https://doi.org/10.1038/s41467-022-30728-3

Greenberg, P., & Ruoff, G. (n.d.). *Malpractice Risks Associated with Electronic Health Records*. Retrieved November 15, 2022, from https://www.rmf.harvard.edu/clinician-resources/article/2017/malpractice-risks-associated-with-electronic-health-records

Kaiser Family Foundation. (n.d.). *KFF Health Tracking Poll - January 2019*. Retrieved November 14, 2022, from https://files.kff.org/attachment/Topline-KFF-Health-Tracking-Poll-January-2019

Kaul, S. D., Kumar Murty, V., & Hatzinakos, D. (2020). Secure and Privacy preserving Biometric based User Authentication with Data Access Control System in the Healthcare Environment. *Proceedings - 2020 International Conference on Cyberworlds, CW 2020*, 249–256. https://doi.org/10.1109/CW49994.2020.00047

Keene, D. J., Lamb, S. E., Mistry, D., Tutton, E., Lall, R., Handley, R., & Willett, K. (2018). *Electronic Health Record Usability Issues and Potential Contribution to Patient Harm*. https://dashboard.healthit.gov

Lee, I. (2022). An analysis of data breaches in the U.S. healthcare industry: diversity, trends, and risk profiling. *Information Security Journal*, *31*(3), 346–358. https://doi.org/10.1080/19393555.2021.2017522

Li, Y., Li, Q., Gao, J., Su, L., Zhao, B., Fan, W., & Han, J. (2016). Conflicts to Harmony: A Framework for Resolving Conflicts in Heterogeneous Data by Truth Discovery. *IEEE Transactions on Knowledge and Data Engineering*, *28*(8), 1986–1999. https://doi.org/10.1109/TKDE.2016.2559481

Low, S., Butler-Henderson, K., Nash, R., & Abrams, K. (2019). Leadership development in health information management (HIM): literature review. *Leadership in Health Services*, *32*(2019), 569–583.

Munana, C., Kiirzinger, A., & Brodie, M. (2019, March 18). *Data Note: Public's Experiences With Electronic Health Records | KFF*. https://www.kff.org/other/poll-finding/data-note-publics-experiences-with-electronic-health-records/

Shaikh, A. K., & Nazir, A. (2020). A novel dynamic approach to identifying suspicious customers in money transactions. *Int. J. Bus. Intell. Data Min.*, *17*(2), 143-158.

Sulkers, H., Tajirian, T., Paterson, J., Mucuceanu, D., Macarthur, T., Strauss, J., Kalia, K., Strudwick, G., & Jankowicz, D. (2018). Improving inpatient mental health medication safety through the process of obtaining HIMSS Stage 7: A case report. *JAMIA Open*, *2*(1), 35–39. https://doi.org/10.1093/jamiaopen/ooy044

The Pew Charitable Trusts, American Medical Association, & Medstar Health. (2018). *Ways to Improve Electronic Health Record Safety Rigorous testing and establishment of voluntary criteria can protect patients Potential drug interaction Warning*.

truenorth. (n.d.). *EHR Usability Issues: Problems with Electronic Health Records*. Truenorth. Retrieved November 14, 2022, from https://www.truenorthitg.com/ehr-usability-issues/